\documentclass{llncs}
\usepackage{amsmath,amssymb,amscd}
\usepackage{hyperref}
\usepackage{rotating} 
\usepackage{graphicx,subfigure}

\newtheorem{conj}{Conjecture}

\newcommand{\secref}[1]{Section~\ref{#1}}

\DeclareMathOperator{\tr}{tr}

\numberwithin{equation}{section}

\newcommand{\Z}{{\mathbb{Z}}}

\newcommand\F{{\mathbb F}}

\newcommand{\gobble}[1]{}
  \newcommand{\rangeref}[2]{%
    \ref{#1}--\afterassignment\gobble\fam 0\ref{#2}%
  }

\makeatletter
\def\imod#1{\allowbreak\mkern5mu({\operator@font mod}\,#1)}
\makeatother

\begin{document}

\title{\vspace{-1in} Stopping time signatures for some algorithms in cryptography}
\author{Percy Deift\thanks{Supported in part by NSF grant DMS-1300965.}\inst{1}, Stephen D. Miller\thanks{Supported in part by NSF grants CNS-1526333 and CNS-1815562.}\inst{2}, and Thomas Trogdon\thanks{Supported in part by NSF grant DMS-1753185.}\inst{3}}

\institute{
Courant Institute of Mathematical Sciences, New York University \\
 \email{deift@cims.nyu.edu}
\and
Department of Mathematics, Rutgers University\\
\email{miller@math.rutgers.edu} \and
Department of Mathematics,
University of California, Irvine\\
 \email{ttrogdon@uci.edu}
  }

\maketitle

\begin{abstract}
We consider the normalized distribution of the overall running times of some cryptographic algorithms, and what information they reveal about the algorithms.  Recent work of Deift, Menon, Olver, Pfrang, and Trogdon has shown that certain numerical algorithms applied to large random matrices
 exhibit a characteristic distribution of running times, which depends {\bf only} on the algorithm but are independent of the choice of probability distributions for the matrices.    Different algorithms often exhibit different running time distributions, and so the histograms for these running time distributions provide a {\it time-signature} for the algorithms, making it possible, in many cases, to distinguish one algorithm from another.
    In this paper we extend this analysis to  cryptographic algorithms,  and present
examples of such algorithms with
   time-signatures that are indistinguishable, and others with
   time-signatures that are clearly distinct.
%

%
\end{abstract}

\section{The phenomenon of running time ``signatures'' in cryptography}\label{sec:intro}

This paper concerns the following issue:
\begin{quote}
What information about an algorithm is revealed just by its distribution of (canonically normalized) running times?
\end{quote}
In other words, do the running times  assign a ``time-signature'' to the algorithm which distinguishes it from other algorithms?\footnote{Of course,  running time can be highly dependent on specific implementations of a given algorithm.  We use the term algorithm to refer a specific implementation.}
Such   time-signatures have  been identified for a variety of algorithms in numerical analysis. In this paper we show that time-signatures also exist for certain   algorithms commonly used in cryptography (such as ones based on finding prime numbers, elliptic curves, or collisions in random walks).

\subsection*{The notion of   ``time-signature''}

We begin with some background on time-signatures from numerical analysis. Starting in 2009, Deift, Menon, and  Pfrang \cite{eigencite4}  considered the running time for the computation of the eigenvalues of random matrices using different algorithms.  In particular, the authors considered real symmetric $n\times n$ matrices $M$ chosen from a probability distribution $\mathcal E$.   They recorded the time $T(M)=T_{\epsilon,n,{\mathcal A},{\mathcal E}}(M)$ to compute the eigenvalues of a matrix $M$ chosen from the distribution $\mathcal E$ to an accuracy $\epsilon$, using  a given  algorithm $\mathcal A$.  Repeating the computation for a large number of matrices $M$ chosen from $\mathcal E$, they plotted the histogram for the normalized times
 \begin{equation}\label{normalizedtimes}
   \tau(M) \ \ = \ \ \tau_{\epsilon,n,{\mathcal A},{\mathcal E}}(M) \ \ := \ \ \frac{T(M)-\langle T\rangle}{\sigma}\,,
 \end{equation}
where $\langle T \rangle$ is the sample average for the times and $\sigma$ is the sample standard deviation.  What they found was that for $\epsilon$ sufficiently small and $n$ sufficiently large, the histogram of $\tau$ depended {\it  only} on the algorithm $\mathcal A$, independent of the choice of $\epsilon$, $n$, and  $\mathcal E$.  Thus the histogram for $\tau$ provided a {\it time-signature} for the algorithm.  We stress that it is not the actual running times which provide the algorithm with a time-signature:~rather, it is the  {\bf fluctuations} of the running times of the algorithm in response to random data that provide the time-signature.

\begin{figure}[tbp]
\centering
\includegraphics[width=.32\linewidth]{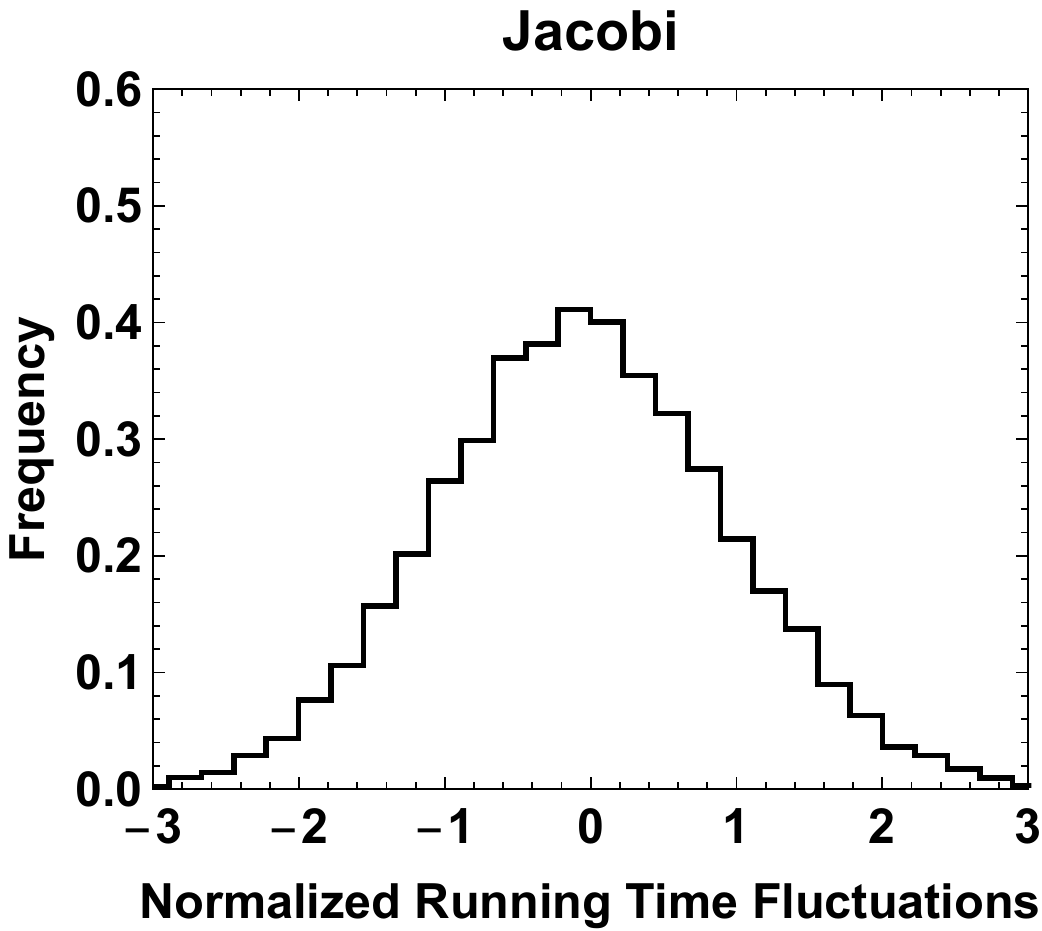}
\includegraphics[width=.32\linewidth]{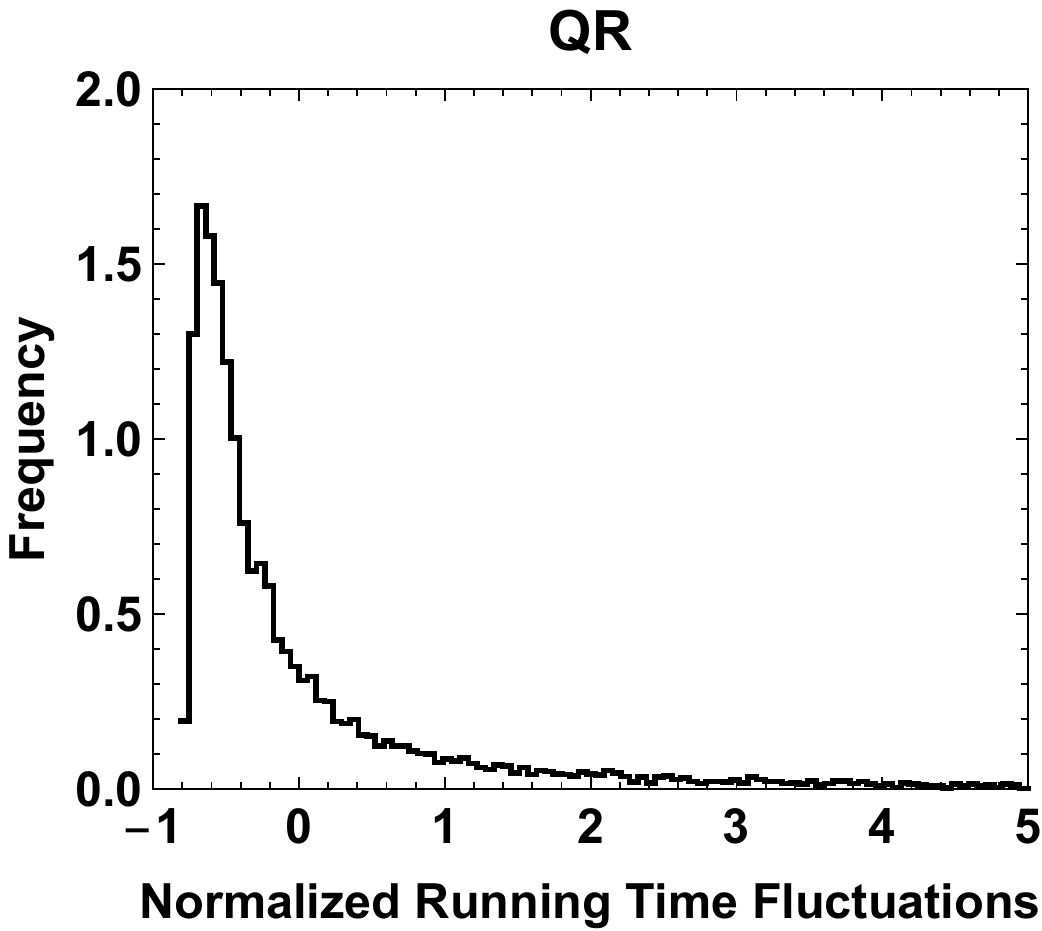}
\includegraphics[width=.32\linewidth]{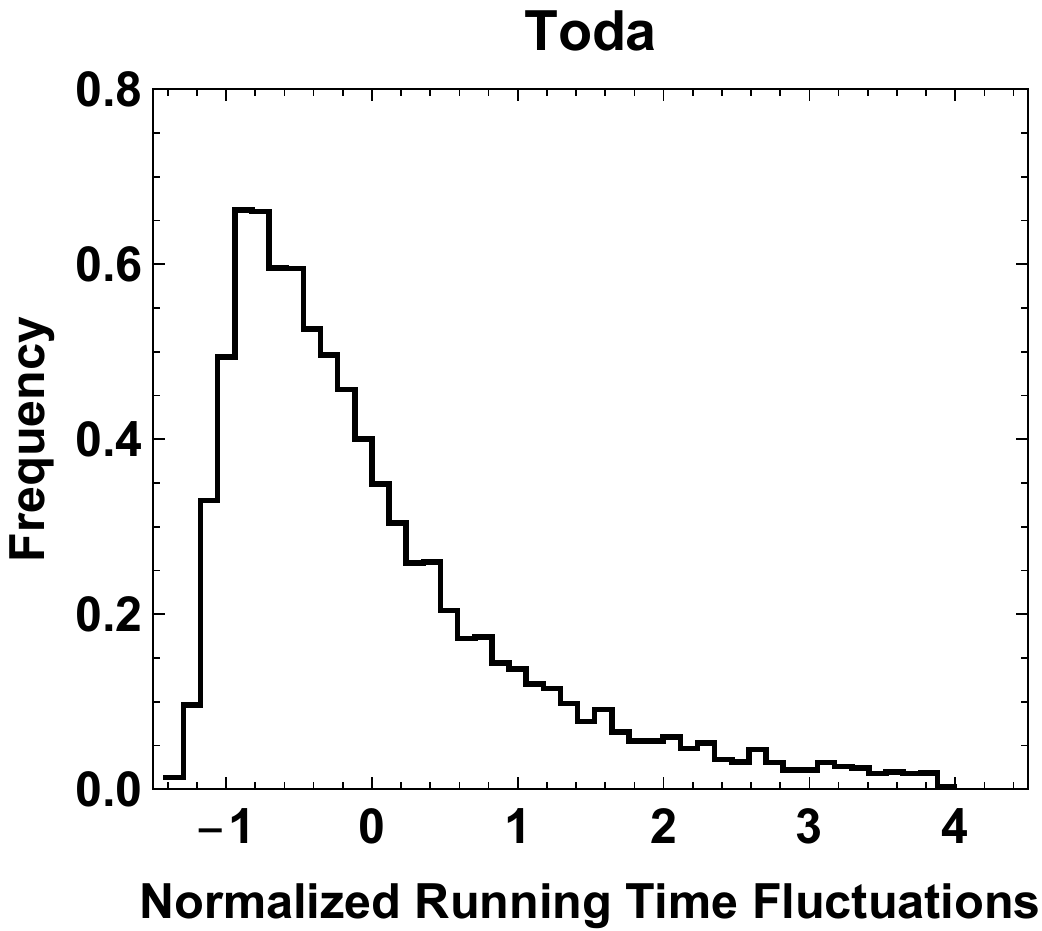}
\caption{The normalized runtime histograms (normalized to mean zero and variance one) for three eigenvalue algorithms (the Jacobi, QR and Toda eigenvalue algorithms) applied to $20 \times 20$ symmetric Gaussian matrices (i.e., the Gaussian Orthgonal Ensemble).  Let $A_k$, $k = 1,2,\ldots,$ be the iterates of an algorithm.  We halt the algorithms when $\tr ((A_k - \mathrm{diag}\, A_k)^2) < \epsilon^2.$ It is clear that the algorithms give distinct histograms.  See \cite{eigencite1} for more detail on these algorithms and computations.}\label{f:sig}
\end{figure}

In later work  with Olver and Trogdon \cite{eigencite1,eigencite3,eigencite2} time-signatures  were found for a wide variety of numerical algorithms.
Moreover, the time-signature was different for different algorithms, making it possible, in particular, to distinguish three specific algorithms (the Jacobi, QR, and Toda  eigenvalue algorithms) from each other.
Said differently, the information carried by the running times {\it alone} is sufficient to distinguish the three eigenvalue algorithms.
As an example, in Figure~\ref{f:sig} we demonstrate that three different algorithms to compute the eigenvalues of a symmetric matrix exhibit three different distributions of runtimes.

In view of the above discussion, two issues are central:

\vspace{.2cm}
 \noindent
 {\bf Problem 1 (Existence)}:
  Suppose an algorithm $\mathcal A$, which  depends on several (large) parameters, acts on random data.  Does the running time of $\mathcal A$ (when normalized to have fixed mean and standard deviation) tend to a limiting distribution as the parameter size tends to infinity in some regime?  If so, does this limit depend on the probability distribution of the data?
\vspace{.2cm}

  \noindent  We refer to this limiting distribution, {\it if} it exists and {\it if} it is independent of the probability distribution of the data, as the {\em time-signature} $\mathcal{S(A)}$ of $\mathcal A$.
\vspace{.2cm}

 \noindent
 {\bf Problem 2 (Uniqueness)}: For a given task of interest, are there known algorithms sharing the same time-signature?
 \vspace{.2cm}

Since the time-signature $\mathcal{S(A)}$ is normalized to have a fixed mean and standard deviation, it measures the overall shape of the distribution of running times, but is insensitive to -- as in the case of the eigenvalue algorithms above -- the actual running times and their dependence on the (large) parameters.

\subsection*{Generating primes}

To demonstrate, at a basic level, the existence of time-signatures in cryptography, we begin with a discussion of the problem of random prime generation.  Indeed, this is a bedrock ingredient of many cryptosystems, such as RSA and Elliptic Curve Cryptography which are studied in this paper.
For $L > 1$, consider a probability distribution $\mu$  on the integers $[L^{N-1},L^N] \cap \mathbb N$.  The prime number theorem tells us that
the density of primes in this interval, roughly $1/(N \log L)$, is not so small  as to make the following prescription   inefficient.  Suppose $X_1,X_2,\ldots$ is a sequence of independent and identically distributed random variables with distribution $\mu$ supported on $[L^{N-1},L^N] \cap \mathbb N$, and let $M > 0$ be an integer.
Then set $j = 1$ and do:
\begin{enumerate}
\item Sample $X_j$.
\item For $\ell = 1,2,\ldots,M$, do:
  \begin{enumerate}
  \item run a Miller-Rabin\footnote{This test first appeared in \cite{rabin}. The Miller--Rabin test investigates whether a given number $X_j$ is a prime, but is only capable of rigorously proving that a number is {\it not} prime.  In particular, it never proves that a number is prime.  If it cannot prove a number is composite, it reports that this number is probably prime -- and  comes with a probability estimate (taken over the random choices the algorithm makes in its execution) that shows it is very unlikely such a probable prime is actually composite.

   In Step 2(a) above one runs Miller-Rabin on a fixed sample $X_j$ repeatedly;
  if it first fails at the $\ell$-th time, $1 \le \ell \le M$, one returns to Step 1 and chooses a new sample $X_{j+1}$. Otherwise, if $X_j$ passes the Miller-Rabin test all $M$ times, the algorithm deems $X_j$ to be prime with high probability (see (\ref{4M})).}
  test on $X_j$;
  \item if $X_j$ fails the Miller-Rabin test set $j = j+1$ and return to 1.
  \end{enumerate}
\item Output $X_j$.
\end{enumerate}

\noindent
Thus this  algorithm allows one to solve the following problem:
  \vspace{.09cm}

\noindent{\bf{Problem 3 (Prime generation):}}
  Produce a random integer $n \in [L^{N-1},L^N]\cap \mathbb N$ such that
  \begin{align*}
    \mathbb P( n \text{ is prime} ) \overset{N \to \infty}{\longrightarrow} 1.
  \end{align*}

%
\noindent
  Indeed, from \cite{rabin}
  \begin{align}\label{4M}
    \mathbb P( n \text{ is composite and } n \text{ passes } M \text{ Miller--Rabin tests} ) \leq 4^{-M},
  \end{align}
 and  it follows that as long as $M$ grows, e.g., $M \sim N^{1/2}$, the problem is solved.

The runtime for the above prime generating algorithm can be modeled by making some simplifying assumptions.  First suppose that the Miller--Rabin test is foolproof on composites, i.e., it immediately and accurately detects all composite numbers\footnote{In practice this happens for random choices with overwhelming probability.
For example, if $L =2$ and $N = 30$ the range $[2^{29},2^{30}]\cap \mathbb N$ contains  26,207,278 primes.  In one experiment, only 361 of the 510,663,634 composite numbers in the range passed at least one Miller--Rabin primality test, a proportion of merely
$6.72415 \times 10^{-7}$.
}.  Then let $T_{\mathrm{MR}}(n;M)$ be the (random) number of seconds it takes to apply $M$ Miller--Rabin tests to an integer $n$. Additionally, suppose that the sampling procedure takes a deterministic number of seconds, $c$.  Define
\begin{align*}
  \tau = \min\{j : X_j \text{ is prime}\}.
\end{align*}
Then the total time for this algorithm can be expressed as
\begin{align}\label{eq:T}
  T = c\tau + \sum_{j=1}^{\tau-1} T_{\mathrm{MR}}(X_j;1) + T_{\mathrm{MR}}(X_\tau;M).
\end{align}
In the case $L = 2$,   the number of primes in the interval $[2^{N-1},2^N]$ is given asymptotically by $\frac{2^{N-1}}{N \log 2}$ according to the prime number theorem.  Therefore, assuming $\mu$ is uniform on $[2^{N-1},2^N] \cap \mathbb Z$, one has
\begin{align*}
    \mathbb E[\tau] = p \sum_{k=1}^\infty k (1-p)^{k-1} = \frac{1}{p}, \  \ \ \text{where} \ \  p = \frac{1}{N \log 2 }(1 + o(1)),
\end{align*}
and for $t > 0$
\begin{align*}
    \mathbb P( \tau \leq t N) = p\sum_{k=1}^{\lfloor t N \rfloor} (1 - p)^{k-1} = 1 - \left(1-\frac{t(1 + o(1))}{ tN  \log 2} \right)^{\lfloor t N \rfloor},
\end{align*}
which converges to $1 - e^{- t/\!\log 2}$ as $N\rightarrow\infty$.
Note the scaling $tN$ is a necessary normalization to obtain a limit which is independent of $N$.

\begin{figure}[tbp]
    \centering
    \includegraphics[width=.9\linewidth]{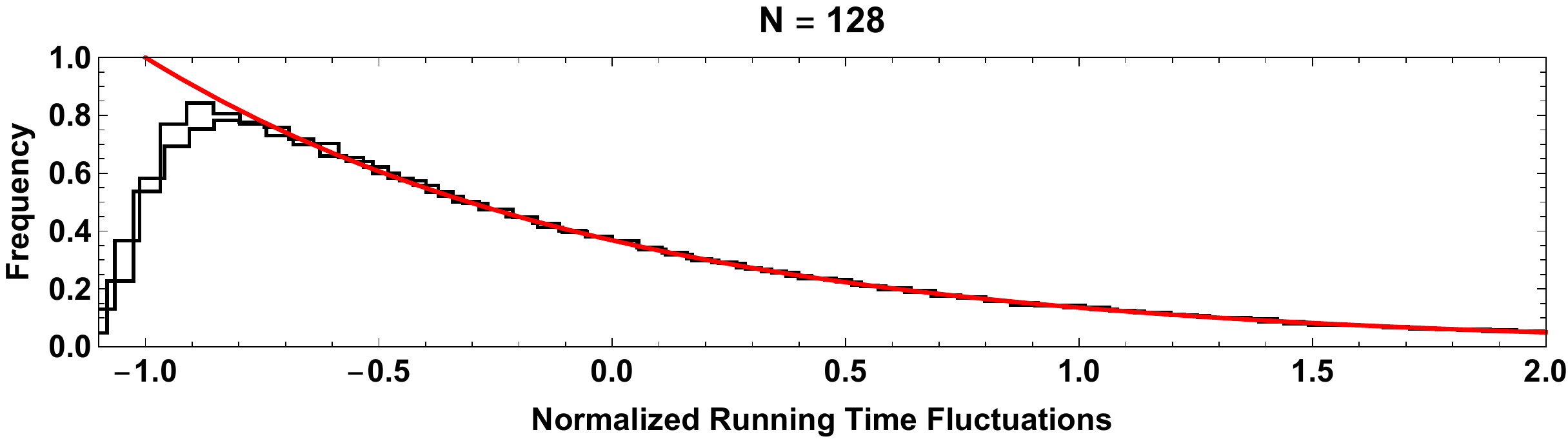}

    \vspace{.2cm}
    \includegraphics[width=.9\linewidth]{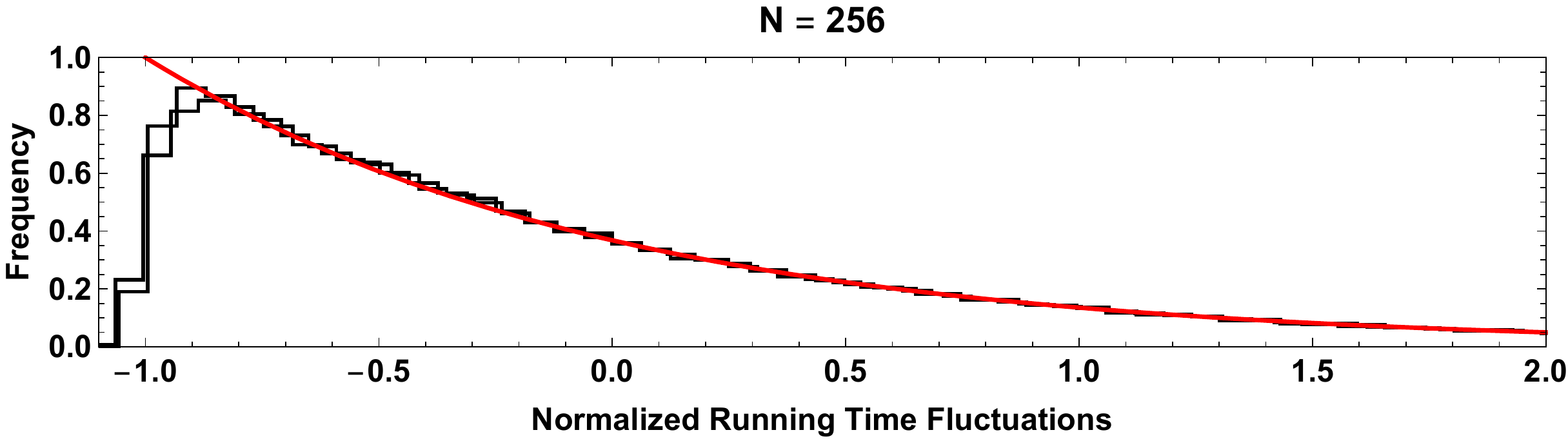}

    \vspace{.2cm} \includegraphics[width=.9\linewidth]{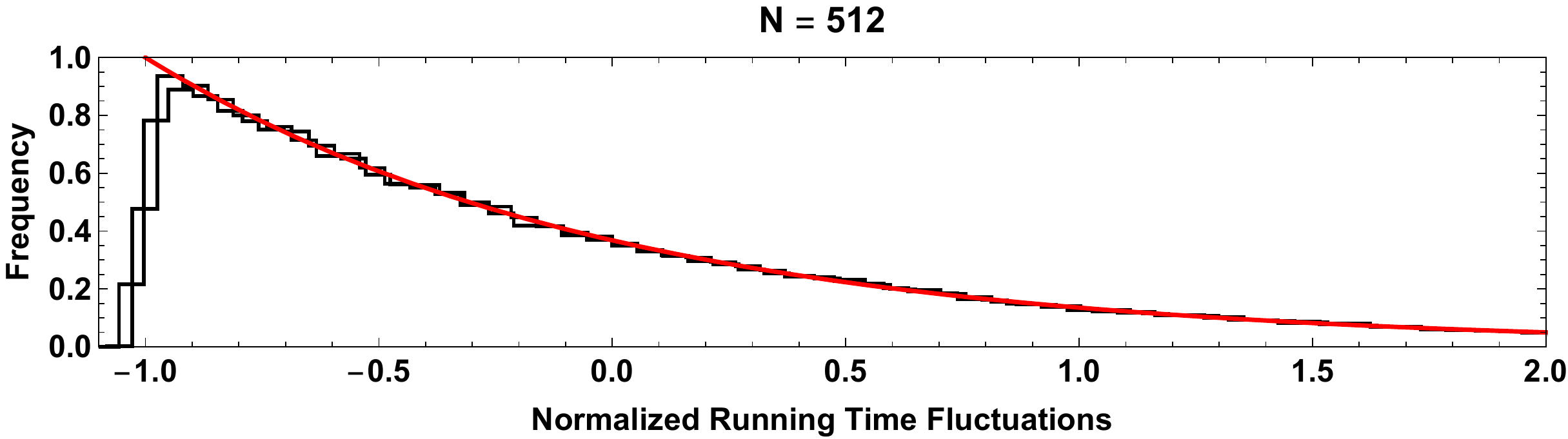}

    \vspace{.2cm} \includegraphics[width=.9\linewidth]{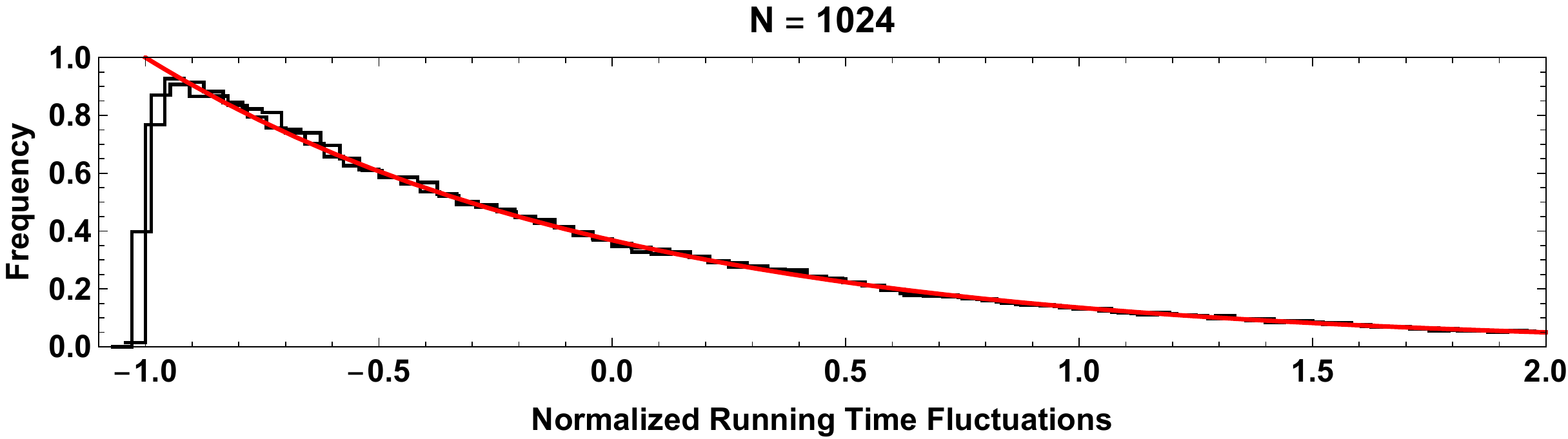}
    \caption{A demonstration of Conjecture~\ref{c:univ} with $M = \lceil \sqrt{\log 2^N} \rceil$. Each panel consists of two histograms, one corresponds to $L = 2$ and $\mu$ being the uniform on $[2^{N-1},2^N]$ and the other corresponds the measure induced on $[2^{N},2^{N+1}]$ by adding two independent copies of random variables with distribution $\mu$.  As $N$ increases, the histograms approach the smooth curve, which is an exponential density with mean zero and variance one.}\label{f:primality}
  \end{figure}

Lastly, suppose for simplicity that $T_{\mathrm {MR}}(X_j;1)$ is a deterministic constant.\footnote{This assumption is an oversimplification in order to motivate the statement of Conjecture 1, which we believe holds without it.} One then expects\footnote{Probably this is true in more generality as long as $\mu$ is not degenerate in some fashion.} that the sum of the first two terms in \eqref{eq:T} is $O(N)$ and converges in distribution to an exponential random variable after rescaling as above. The last term in (\ref{eq:T}) is clearly $O(M)$, and as long as $M \ll N$  we can use this analysis to conjecture a limiting distribution for $T$:

  \begin{conj}[Universality]\label{c:univ}
    Suppose $N \gg M$. Then for a wide class of distributions $\mu$
    \begin{align*}
      \mathbb P \left( \frac{ T - \mathbb E[T]}{ \sqrt{\mathrm{Var}(T)}} \leq t \right) = 1 - e^{-t-1}.
    \end{align*}
  \end{conj}
  \noindent We demonstrate the conjecture with numerical examples in Figure~\ref{f:primality}.

\subsection*{Time-signatures for RSA and Elliptic Curve Cryptography}

Running times have long played a role in an indirect form of cryptanalysis known as ``side-channel'' attacks, where instead of directly tackling the underlying computation one also leverages   indirect information about the device   performing it.
In so-called ``white-box'' cryptographic contexts,   the inner workings of the algorithm are kept secret and
 discovering the identity of the algorithm must itself be part of the attack.  Here  ``side-channel'' information such as  power consumption and running time  can be helpful.  An example is   the {\sc KeeLoq} remote   entry system, which is used on many automobile key chain fobs to lock/unlock vehicles:~the attack in \cite{keeloq} shows how to easily clone  fobs from power measurements, with only partial access to algorithmic details.   This also provides    motivation to study the mathematical basis of using running time   to determine the identity of an algorithm.
  Recall that we are not concerned about the absolute running times, but instead the shape of their normalized distribution (e.g., which should be independent of hardware changes).
%
%

As an example of time-signatures for cryptographic algorithms, consider  the dominant public-key cryptosystems in use today:~RSA and Elliptic Curve Cryptography (see  Appendix~\ref{sec:crypto} for  details).  Generating keys for each algorithm involves random choices of integers of  varying sizes, and the way in which these random choices are made can vary.   One then verifies numerically that as the key size grows, Problem 1 has a positive answer for both algorithms.  That is, each algorithm has a time-signature, which is  independent of the way the random choices are made.
\begin{figure}[h]
  \centering
  {\includegraphics[width=\linewidth]{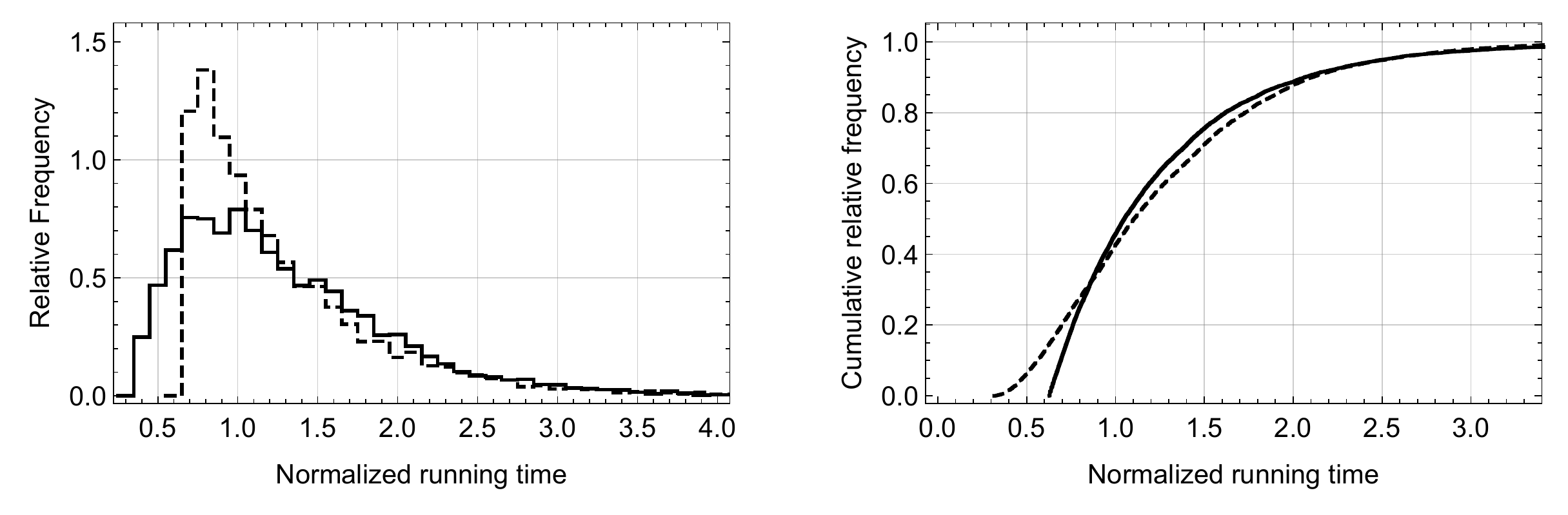}
  \caption{The   distribution   of time-signatures for RSA key generation (dashed curve) vs. Elliptic Curve Cryptography key generation (solid curve).  The plot on the left shows histograms, i.e., probability distribution functions,  while the plot on the right shows their cumulative distribution functions.  Each curve reflects actual running times using the implementation described in Appendix~\ref{sec:sub:figures}.  In contrast with   examples later in this paper, these  time-signatures can be easily distinguished. \label{fig:ECCvsRSAnewa}  }}
\end{figure}
Figure~\ref{fig:ECCvsRSAnewa} shows a plot of the time-signatures for both algorithms, which are visibly different.   Thus we see that
  Problem 2, in the context of distinguishing RSA key generation from elliptic curve key generation, has a negative answer (as was the case for the eigenvalue algorithms mentioned above).

This shows that certain cryptographic algorithms have time-signatures and that, at least in concept,
time-signatures can distinguish two different key generation algorithms.  Our next example studies two other cryptographic algorithms (mainly used in attacking cryptosystems but sometimes used in building them as well) that have a common time-signature.



\subsection*{Random walks and discrete logarithms}

    We next turn to the Discrete Logarithm Problem (see (\ref{DLOG})) and shall see that two algorithms to solve it,  Algorithms A.2.1 and A.2.2 in Appendix~\ref{sec:sub:dlog}, have   identical time-signatures.
    These algorithms are based on finding repetitions in sequences of numbers;~these sequences are described in Processes 1 and 2 just below, and their connection with discrete logarithms is described in
    Appendix~\ref{sec:sub:dlog}.
      Processes 1 and 2 involve collision times of random walks on a set of $N$ elements, which is taken to be $\Z/N\Z$ for concreteness.
      It turns out that both algorithms share the {\it same}  time-signature:~the Rayleigh   probability distribution on $[0,\infty)$,
\begin{equation}\label{pray}
  p_{\text{Ray}}(x) \ \ = \ \ xe^{-x^2/2}\ , \ \ \ \int_{0}^x p_{\text{Ray}}(t)\,dt \ \ = \ \ 1\,-\,e^{-x^2/2}\,.
\end{equation}
    For our purposes it is more natural to normalize the running times by
     \begin{equation}\label{normalizedtimesRay}
   {\mathcal N}(T) \ \ := \ \ \sqrt{2-\frac{\pi}{2}}\,\frac{T -\langle T\rangle}{\sigma}+\sqrt{\frac{\pi}{2}}
 \end{equation}
     instead of (\ref{normalizedtimes}) in order to match the mean and standard deviation of (\ref{pray}).
  \\

\begin{figure}
   \includegraphics[width=.74\linewidth]{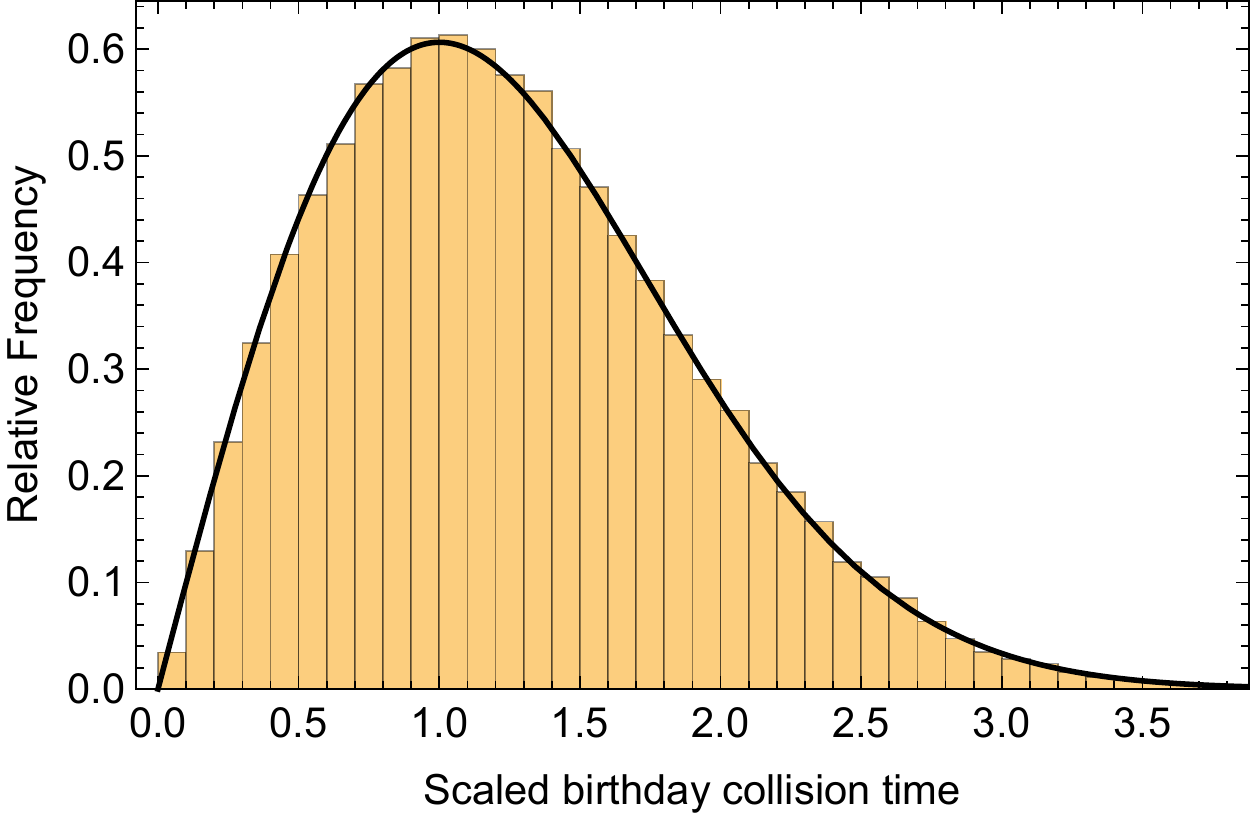}
   \centering
\caption{A demonstration that the distribution ${\mathcal N}(B_N)$ from Process 1 converges to the Rayleigh distribution (solid line).    Here $N=40,\!009$ (the first prime larger than 40,000) and the histogram was generated from $100,\!000$ samples.
The same plot   with $B_N$ replaced by collision times for the Pollard $\rho$ walk from Process 2, with ensemble $\cal E_\rho$ as in Section~\ref{sec:prho}\label{fig:Rayvshist}, looks identical.}
\end{figure}

\paragraph{\bf Process 1: Uniform random walks and the ``birthday problem''.}    This is the simplest random walk  on   $\Z/N\Z$, in which each element of a sequence $(x_n)_{n\ge 1}$ is chosen uniformly and independently at random.  The famous ``birthday paradox'' asserts that the first collision time $B_N$ of this random walk (the smallest integer $n$ such that $x_k=x_n$ for some $1\le k<n$) is roughly $N^{1/2}$ in size.  More precisely,    $N^{-1/2}B_N$ follows the Rayleigh distribution (\ref{pray})  for $N$ large (see, for example, \cite{CP}).   In particular, $B_N$ has mean which grows as  $\sqrt{\frac{\pi}{2}N}$ for $N$ large.  A   histogram for $B_N$, normalized as in (\ref{normalizedtimesRay}), is given in Figure~\ref{fig:Rayvshist}.

If one changes the uniform distribution of each $x_n\in \Z/N\Z$ to a nonuniform distribution on $\Z/N\Z$, the random walk is  biased towards a smaller subset, which reduces the expected value of $B_N$.  Nevertheless, the     appearance  of the Rayleigh distribution persists:~Camarri and Pitman \cite{CP} have given general conditions for $\mu N^{-1/2}B_N$ to tend to the Rayleigh distribution as $N\rightarrow\infty$, with a constant  $\mu$ given by an explicit formula.  In particular, the birthday problem indeed has an associated time-signature, namely (\ref{pray}).\\

The next type of walk has correlations between the $x_n$'s.  The particular example we now describe first appeared in the Pollard $\rho$ Algorithm A.2.3 for computing discrete logarithms on cyclic groups of order $N$, where it was originally devised as a pseudorandom walk \cite{Pollard} (see \secref{sec:prho} and Appendix~\ref{sec:crypto}   for more details).  More generally it is representative of a   class of random walks on certain directed  graphs that we shall also see    share the same normalized collision time histograms (see Section~\ref{sec:expander}).  In particular, algorithms halting upon collisions of such random walks have a time-signature:~moreover, they have the same time-signature as for the birthday problem, viz.,
the   Rayleigh distribution (\ref{pray}), even though the (unnormalized) running times are noticeable distinct.

\paragraph{\bf Process 2: Pollard $\rho$ walk.}
In contrast to Process 1, the walk moves in at most three possible directions at any stage.
The Pollard $\rho$ walk is determined by two objects, an element $h\in \Z/N\Z$ and a decomposition
of $\Z/N\Z$ into three disjoint sets, $S_{1,\text{key}}$, $S_{2,\text{key}}$, and $S_{3,\text{key}}$ of roughly equal size.
  Starting from some point $x_1\in \Z/N\Z$, we define a sequence $(x_n)_{n\ge 1}$ via the iteration  $x_{n+1}=f_{{\text key},h}(x_n)$, where the iterating function  $f_{{\text key},h}:\Z/N\Z\rightarrow \Z/N\Z$ is defined by
  \begin{equation}\label{fkey}
 f_{{\text key},h}(x) \ \ = \ \ \left\{
                                 \begin{array}{ll}
                                   2x, & x\in S_{1,\text{key}}\,; \\
                                   x+1, & x\in S_{2,\text{key}}\,; \\
                                   x+h, & x\in S_{3,\text{key}}\,.
                                 \end{array}
                               \right.
  \end{equation}
  Randomness can enter into the walk in three ways:~1) by choosing $h$ uniformly at random from $\Z/N\Z$; 2) by constructing the decomposition using a random prescription depending on a ``key'' (see Section~\ref{sec:prho} for more details); and 3) by choosing the initial point $x_1\in \Z/N\Z$ randomly.  Note that once the choices 1), 2), and 3) have been made, the sequence $x_1,x_2,\ldots$ is completely deterministic.
   The walk is most interesting when $N$ is prime, in which case it was proved to have collisions in time $O(N^{1/2})$ with high probability with respect to the choice of randomness for the decomposition $\Z/N\Z=S_{1,\text{key}}\sqcup S_{2,\text{key}}\sqcup S_{3,\text{key}}$   (see \cite{KMPT,Mo,MV1,MV2}).

     We have performed several computations of the collision time distribution, as follows:~for  several different choices of $N\gg 1$  and several different choices of  $h,x_1\in \Z/N\Z$, we record the collision time of the walk for various
     decompositions $\Z/N\Z=S_{1,\text{key}}\sqcup S_{2,\text{key}}\sqcup S_{3,\text{key}}$ chosen at random (see Section~\ref{sec:prho}).  What emerges from these computations is a  histogram for the collision times, normalized as in (\ref{normalizedtimesRay}), which looks to be identical to the Rayleigh distribution shown in Figure~\ref{fig:Rayvshist}.     In other words, the Pollard $\rho$ walk indeed has an associated time-signature, and the time-signature is the same as for the birthday problem.
   \\

The time-signature of the algorithm does not depend on the raw running time, but rather  {\it reflects the response of the algorithm to randomness} in a canonical way.
For example,
the mean collision time for the Pollard $\rho$ walk on $\Z/N\Z$ is expected to be $\approx 1.6N^{1/2}$ \cite{T}, which is significantly higher than the mean collision time of $\sqrt{\frac \pi 2}N^{1/2}\approx 1.3N^{1/2}$ for the ``birthday problem'' random walk considered above.
  It is perhaps surprising that the normalized histograms of collision times are nevertheless insensitive to this fundamental distinction.

\begin{figure}[h]
  \centering
  {\includegraphics[width=.75\linewidth]{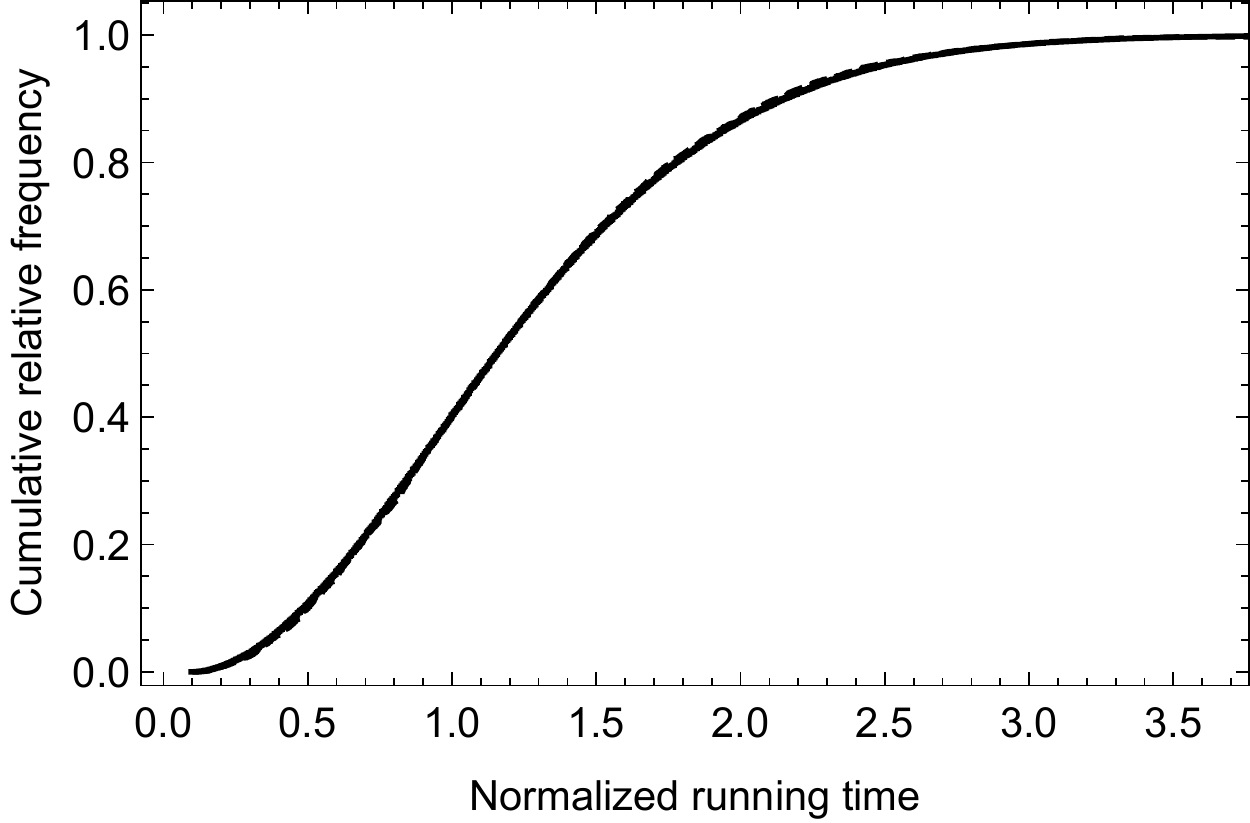}
  \caption{\label{fig:ECCvsRSAnewb}   A comparison of time-signatures of the same algorithm, the Pollard $\rho$ Algorithm A.2.3 for discrete logarithms on elliptic curves, with running time measured two different ways.  The solid curve shows the time-signature as measured by clock time, while the dashed curve shows the time-signature as measured by the number of iterative steps.  The closeness of these   cumulative distribution functions of these time-signatures is apparent.  See Appendix~\ref{sec:sub:figures} for more details. }}
\end{figure}

Our comparisons of RSA and elliptic curve key generation in  Figure~\ref{fig:ECCvsRSAnewa} involved actual CPU running time, i.e., clock time.  By contrast, our discussion of collision times instead counted the number of iterative steps as a proxy for running time.  This is done as a matter of mathematical convenience.
 For example, in particular, the clock times and the running times for one of these algorithms, viz., the Pollard $\rho$ Algorithm A.2.3, are essentially indistinguishable, as can be seen in Figure~\ref{fig:ECCvsRSAnewb} (see also Appendix~\ref{sec:sub:figures} for further discussion). 

   In \secref{sec:prho} we go into more detail about  the Pollard $\rho$ algorithm, and highlight an observation concerning the distribution of its running times.  In  \secref{sec:expander} we address more purely mathematical issues and
   consider random walks on graphs, through which our observations about the Pollard $\rho$ algorithm can be seen as a special case of a more-general phenomenon.  Concluding remarks are made in \secref{sec:conclusion}.  For the benefit of mathematicians, in  Appendix~\ref{sec:crypto} we have included   cryptographic background and   details about the cryptographic constructions just described, along with more details of our experiments.


\section{Pollard $\rho$ collision times}\label{sec:prho}

Recall that the Pollard $\rho$ walk involves a random decomposition of $\Z/N\Z$ as $S_{1,\text{key}}\sqcup S_{2,\text{key}}\sqcup S_{3,\text{key}}$.  The iterating function (\ref{fkey}) depends on the decomposition, but in practice  it is prohibitive to store the decompositions themselves.  In order to make the algorithm efficient, one instead chooses the decompositions based on the value of some fixed, keyed   hash function $H_{\text{key}}(x)\pmod 3$.\footnote{The hash function is a function from $\Z/N\Z\rightarrow \Z$ with the important property that it can be  evaluated quickly.  A {\it keyed hash function} is  a collection of hash functions, one for each choice of ``key'' (which can be represented, for example, by an integer).  In this paper the keys are chosen randomly.
Of course all analyses (including ours) assume the hash function outputs are independent and equidistributed$\pmod 3$.}   Thus a random choice of key determines a random decomposition  and (\ref{fkey}) can be rewritten as
  \begin{equation}\label{fkeyHash}
 f_{{\text{key}},h}(x) \ \ = \ \ \left\{
                                 \begin{array}{ll}
                                   2x, & H_{\text{key}}(x)\,\equiv\, 0\pmod 3\,; \\
                                   x+1, & H_{\text{key}}(x)\,\equiv\, 1\pmod 3\,; \\
                                   x+h, & H_{\text{key}}(x)\,\equiv\, 2\pmod 3\,.
                                 \end{array}
                               \right.
  \end{equation}
  Observe that once $h$, the key, and the starting point $x_1$ have been chosen, the algorithm is deterministic.
  From the point of view of the experiments in this paper, the underlying ensemble $\cal E_\rho$ for the Pollard $\rho$ walk consists of probability distribution functions for the choices of $x_1$, $h$, and the key.  In our experiments, we have demonstrated universality for the collision time $s_N$ with respect to various such ensembles.  More precisely, our calculations demonstrate the following
%
\\

\noindent
{\bf Observation 1 (Pollard $\rho$ walk universality):} Let $N$ be a prime, $c>0$ a  fixed constant, and $\overrightarrow{\mathbf x}$ a list of $cN^{1/2}+O(1)$ samples of  $s_{N}$.  Then for a wide class of ensembles $\cal E_\rho$, ${\mathcal N}(\overrightarrow{\mathbf x})$ converges to the Rayleigh distribution (\ref{pray})  as $N\rightarrow \infty$.\\

\noindent
The list is taken to have size on the order of $N^{1/2}$ which is large enough to generate meaningful samples, but small enough to be practical for experiments.
This universality was tested extensively, in particular using Kolmogorov-Smirnov statistics to test the rate of convergence to the Rayleigh distribution.
In order to generate a data set large enough to obtain valid statistical conclusions, but not so large that the  $O(\sqrt{N})$ running time becomes prohibitive, we took
 the value of $N$ in our experiments to be on the order of $10^5$ (e.g., as in Figure~\ref{fig:Rayvshist}).
 The ensembles $\cal E_\rho$ we studied were appropriately regular in their distributions of $x_1$ and $h$, in order to avoid artificially small collision times.  In particular, we do not allow distributions which assign high probability to $x_1=0$, for then $x_2=0$ with probability at least 1/3 and   there is frequently a collision at the first possible opportunity. Similar caveats apply also in the simpler situation of the birthday problem.\\

\noindent
{\bf Collision statistics for the   Pollard $\rho$ Algorithm A.2.3.}   We have just discussed the collision times of the Pollard $\rho$ {\it walk}, which differs from the running time of the actual Pollard $\rho$ algorithm since the latter uses Floyd's cycle detection method described in Appendix~\ref{sec:sub:dlog}.  The Pollard $\rho$ algorithm does not necessarily detect the first collision, but instead waits for the algorithm to enter into a loop.  In particular,
 the number of iterative steps taken to find a collision in the Pollard $\rho$ Algorithm A.2.3 is at least that of Algorithm A.2.2.

 We performed a number of experiments with the Pollard $\rho$ Algorithm A.2.3 itself, and found that the  normalized histograms of their collision times also converge to a limiting time-signature.  Intriguingly, this time-signature is
   very close to the Rayleigh distribution limit that appears in Observation 1, but distinct from it:~their cumulative  distributions  appear  to be within 1\% of each other.
    This closeness is striking, and appears to reflect the robustness of the universality of the Rayleigh distribution (\ref{pray}).

\section{Collision times of random walks on graphs}\label{sec:expander}

In this section we make some speculations about time-signatures of collision times of random walks on graphs.

Recall the discussion of the Pollard $\rho$ walk from (\ref{fkey}).
 Once  the decomposition $\Z/N\Z=S_{1,\text{key}}\sqcup S_{2,\text{key}} \sqcup S_{3,\text{key}}$, the value of $h$, and the starting point $x_1$ have been fixed, the walk is completely deterministic   (cf.~the comments just after (\ref{fkeyHash})).  In particular, the value of each iterate $x_{n+1}$ is completely determined by $x_n$ according to the membership of $x_n$ in $S_{i,\text{key}}$, $i\in\{1,2,3\}$.  As noted before, this is why the algorithm enters into a loop after the first collision.
Part of the randomness in the iteration arises from    the {\it a priori} choice of  the   decomposition    $\Z/N\Z=S_{1,\text{key}}\sqcup S_{2,\text{key}} \sqcup S_{3,\text{key}}$.
The set of such decompositions is in bijective correspondence with the set of all functions from $\Z/N\Z$ to $\{1,2,3\}$, and has cardinality $3^N$.  We let ${\mathcal E}_{\text{decomp}}$ denote the ensemble of all such decompositions, each given equal weight $3^{-N}$.  Note that this differs from the situation in Section~\ref{sec:prho}, where for practical reasons the   choice of decomposition  was delegated to the choice of a key for the keyed Hash function $H_{\text{key}}$.   Since
  not all decompositions   necessarily arise from keys in this way,   for the theoretical considerations of this section we  now  consider ${\mathcal E}_{\text{decomp}}$  directly.

Recall that a directed graph is a set of vertices together with some edges connecting specified pairs of vertices with a specified orientation.
The Pollard $\rho$ iteration can be interpreted as a walk on a  directed graph.
The Pollard $\rho$ graph
$\Gamma_{N,h}$ is defined as the directed  graph on $\Z/N\Z$ having directed edges from each $x\in \Z/N\Z$ to $x+1$, $x+h$, and $2x\in \Z/N\Z$.  Note that these quantities are not always distinct -- the graph contains a small number of self-loops and multiple edges.  Once the decomposition $\Z/N\Z=S_{1,\text{key}}\sqcup S_{2,\text{key}} \sqcup S_{3,\text{key}}$ has been fixed, the iterating function $f$  steers the walk deterministically from each vertex to the next according to (\ref{fkey}).

It is mathematically natural to study {\it uniform random} walks on $\Gamma_{N,h}$, $h$ fixed, meaning that at each stage one obtains the next iterate by choosing  one of the three directed edges with equal probability:~thus the value of $x_{n+1}$ can be either $x_n+1$, $x_n+h$, or $2x_n\in\Z/N\Z$ -- each with probability 1/3, independently of $n$ or the value of $x_n$.  Such a walk will not necessarily enter into a loop after the first collision, because it will generally move  out of the collision site differently the second time around.

It was observed in \cite{MV1} that the collision time statistics of the uniform random walk on $\Gamma_{N,h}$ for fixed $h,x_1\in \Z/N\Z$, are identical to those of the Pollard $\rho$ walk with the same values of $h$ and $x_1$, when the decomposition   $\Z/N\Z=S_{1,\text{key}}\sqcup S_{2,\text{key}} \sqcup S_{3,\text{key}}$ for the Pollard $\rho$ walk is sampled from $\mathcal E_{\text{decomp}}$.  This is because until a collision occurs, it is irrelevant whether the next iterate was chosen from a rule based on predetermined random choices, or if it was instead chosen randomly at the time.
Note that since $h$ and $x_1$ are not randomized at this point, one cannot expect this common distribution to converge to the Rayleigh distribution as $N\rightarrow\infty$, as  in Observation~1.


Thus Observation 1 can be reformulated as saying that the collision time of the random walk on the Pollard $\rho$ graph, for random initial data $h$ and $x_1$, has a time-signature.
It is   intriguing to consider whether or not there are time-signatures for random walks on other types of  graphs, and if so, whether they also follow the Rayleigh distribution.  Notice, for example, that one can view the birthday problem   Process 1 in the introduction as a random walk on the {\it complete} directed graph on $\Z/N\Z$, and indeed its normalized collision times follow the Rayleigh distribution. We performed extensive numerical experiments with the  graph\footnote{We chose to study the graph  $\Gamma_N$ for two reasons:~1) its expansion and eigenvalue-separation properties are similar to those of the Pollard $\rho$ graph, and 2)  its speed of operation.  The latter is due to the fact that   the underlying computational steps describing the edges of $\Gamma_N$ (adding and subtracting 1, and multiplying by 3) are among the simplest and fastest arithmetic operations, and are hence desirable for use in fast stream ciphers (e.g.,  {\sc mv3} \cite{KMMV}).} $\Gamma_N$, defined for primes $N>3$ as the  2-valent directed graph on $\Z/N\Z$ with edges of the form
$$x\rightarrow 3(x\pm 1)\in \Z/N\Z$$
We found again that the normalized collision times follow the Rayleigh distribution (even at the fine level of Kolmogorov-Smirnov statistics).  More precisely, our calculations demonstrate\\

\noindent {\bf Observation 2 (Universality of collision times on    graph $\Gamma_N$):} Let $S_N$ denote the collision time of the uniform random walk on $\Gamma_N$, randomized over uniformly-chosen choices of starting point.
Let $c>0$ be a fixed constant and $\overrightarrow{\mathbf x}$ a list of $cN^{1/2}+O(1)$ samples of  $S_N$.  Then  ${\mathcal N}(\overrightarrow{\mathbf x})$ converges to the Rayleigh distribution (\ref{pray}) as $N\rightarrow\infty$.\\

It is a great challenge to turn Observations 1 and 2 into theorems.  We {\bf conjecture} that   universality phenomena similar to Observation 2 hold for a broad class of random walks on graphs.

\section{Conclusion} \label{sec:conclusion}

We have shown that the notion of time-signature, previously studied in the numerical analysis literature, can be applied to certain cryptographic algorithms.  The shapes of their associated histograms can distinguish between different algorithms, but not always:~several families of random walk-based algorithms have indistinguishable time-signatures.  The connection described in \secref{sec:expander} to the  collision times of random walks on graphs raises interesting mathematical questions  in probability theory.

Future, more-detailed work may create a tool to detect the identity of a hidden algorithm based on  features of its performance such as the distribution of  running times, which could potentially help expose the identity (or non-identity) of a secret implementation of a cryptosystem, or   information about  secret information in a known system.  Time-signatures could also be generalized to other side-channel information, such as power consumption.

\paragraph{\bf Acknowledgements:} It is a pleasure to thank Adi Akavia, Daniel J. Bernstein,  Orr Dunkleman, Nicholas Genise,  Nathan Keller, Neal Koblitz, Thomas Kriecherbauer, Ilya Mironov, Adi Shamir, Ramarathnam Venka\-tesan, and Moti Yung for their helpful comments and discussions.

\appendix

\section{Cryptographic Background}\label{sec:crypto}

In this appendix we provide some  background for mathematicians who may not be familiar with public key cryptography,
 along with more details on the plots in Figures~\ref{fig:ECCvsRSAnewa} and \ref{fig:ECCvsRSAnewb}.

\subsection{Three classic  public key cryptosystems}\label{sec:sub:DHRSA}

\paragraph{\bf The Diffie-Hellman Key Exchange.}  This was the first method for two parties, Alice and Bob, to agree upon a common  secret
 --
 even if they have never previously communicated,  and even if malicious actors eavesdrop on all their communications.  First Alice and Bob   agree on a finite cyclic group $G$ and a generator $g$ of $G$.\footnote{In their original application \cite{DH}, Diffie and Hellman chose $G=(\Z/p\Z)^*$ for some large prime $p$.}  Alice chooses a secret integer $x$, and sends $g^x\in G$ to Bob.  Meanwhile, Bob chooses a secret integer $y$ and sends $g^y\in G$ to Alice.  Each party can then compute the shared secret $g^{xy}=(g^x)^y=(g^y)^x\in G$.   The secret would be  found immediately by an attacker who can solve the
\begin{multline}\label{DLOG}
  \text{Discrete Logarithm Problem (DLOG): }
  \\
   \text{given $h\in G$, find some integer $z$ such that $g^z=h$\,,}
\end{multline}
that is, who can invert the map $a\mapsto g^a$.  Such an attacker who has access to the public information $g^x$ (or $g^y$) could then compute the secret integer  $x$ (or $y$).  In practice, the size of $G$ is extremely large (e.g., order $2^{160}$), making naive approaches (such as a brute-force search for $x$ among all $\#G$ equivalent possibilities) prohibitive.  Algorithms A.2.1, A.2.2, and A.2.3, we shall see, are better.
\\

Once two parties agree on a shared secret, they can encrypt messages via more traditional methods (as a naive example, they can add this shared secret  to a numerical message, and then decrypt by subtracting).

\paragraph{\bf The Rivest-Shamir-Adleman (RSA) cryptosystem.} The RSA algorithm is a  method for one party, Charlie, to directly send an encrypted message to another party, Debra -- without   needing to agree on a secret key, or even  to communicate at all beforehand.  First Debra selects two  large prime numbers, $p\neq q$, and a pair of integers $e$ and $d$  satisfying $ed\equiv1\pmod {\phi(n)}$, where  $n=pq$ and $\phi(n)=(p-1)(q-1)$ is the order of the multiplicative group $(\Z/n\Z)^*:=\{a\in \Z/n\Z\,|\, \gcd(a,n)=1\}$.  Debra publishes the pair $(n,e)$, her {\it RSA key},
but keeps $p$, $q$, and $d$ secret.  To send a message $m$, thought of as an element of $(\Z/n\Z)^*$, Charlie exponentiates $m$ to the power $e$ (the encryption exponent) modulo $n$, and then sends $x\equiv m^e\pmod n$ to Debra, who can recover $m$ as $x^d\pmod n$ using the fact that $x^d\equiv m^{ed}\equiv m \pmod n$ (since $ed-1$ is a multiple of  $\phi(n)$, the order of the group $(\Z/n\Z)^*$).  The system would  be immediately broken if an attacker can factor $n$ into $p$ and $q$, in which case $d$ could be computed as the modular inverse of $e$ modulo $(p-1)(q-1)$.

\paragraph{\bf Elliptic Curve Cryptography (ECC).}  This is the Diffie-Hellman key-exchange applied to the group $G$ of points of an elliptic curve $E$ over a finite field $\mathbb{F}$.  Typically one considers elliptic curves of the form $E:y^2=x^3+ax+b$ with coefficients $a,b\in \F$.  The solutions
$(x,y)\in \F\times \F$ to $E$ along with an additional ``point at infinity''   form an abelian group    under an addition law coming from the geometry of the curve.  To keep with the description of the Diffie-Hellman scheme presented above, we shall write the group operation in multiplicative form; that is, $g^x$ denotes adding the point $g$ to itself $x$ times.

\subsection{Cryptographic algorithms based on the random walks in Section~\ref{sec:intro}}\label{sec:sub:dlog}

We will next explain how both
  Process 1  (the birthday problem random walk) and Process 2 (the Pollard $\rho$ walk) in the introduction can be applied to solve the Discrete Logarithm Problem (DLOG) (\ref{DLOG}).  Afterwards, we will describe how these processes also appear in cryptosystems themselves (i.e., not just in attacks on cryptosystems).

   The algorithms described here are so-called ``black box'' attacks,
meaning that they are designed to work on any finite cycle group -- in particular, they do not leverage any features of the actual realization of the group $G$.   Indeed, much faster attacks are known for certain special realizations of groups.  The attacks below are essentially sharp in their running time (except where otherwise noted) for general groups \cite{KMPT,MV2,shoup}.  It is worth commenting that the Pollard $\rho$ Algorithm A.2.3 is one of the few examples of an algorithm whose running time is rigorously understood; in general most analysis of running times is only heuristic.\footnote{
As an extreme example, the presently most-powerful integer factorization algorithm (the ``Number Field Sieve'') is not even provably known to  terminate.  More generally, the halting problem (determining whether or not an arbitrary program will terminate) is undecidable.}  With high probability the running time is bounded above and below by constant multiplies of $\sqrt{\# G}$, which is a dramatic improvement over the   $O(\#G)$ running time for a brute-force search.

\paragraph{\bf Algorithm A.2.1:~DLOGs using the birthday problem random walk.}
Given $h\in G$, we must find some integer $x$ such that $g^x=h$.  Starting with $x_1=h$, continue with the iteration $x_{n+1}=x_n^{r(x_n)}g^{s(x_n)}$, where $r(x_n)$ and $s(x_n)$ are  integers  chosen uniformly at random from the  interval $[1,\#G]$, and which depend only on $x_n$ (but not the value of $n$ itself).  At each stage, $x_n$ has the form $h^{a_n}g^{b_n}=g^{a_n x+b_n}$, where $a_n$ and $b_n$ are known integers.  Since the exponent  $s_n$ is chosen  uniformly at  random,   $x_{n+1}$ is uniformly distributed at random, and hence one expects based on the birthday paradox a collision $x_m=x_k$ for some $1<k<m=O(\sqrt{\# G})$.  Comparing exponents results in the linear relation $a_m x+b_m\equiv a_kx+b_k\pmod{\# G}$; this can be solved as long as $a_m-a_k$ is coprime to $\# G$, which happens with high probability.

      \begin{figure}[h]
     \centering
     \includegraphics[width=.6\linewidth]{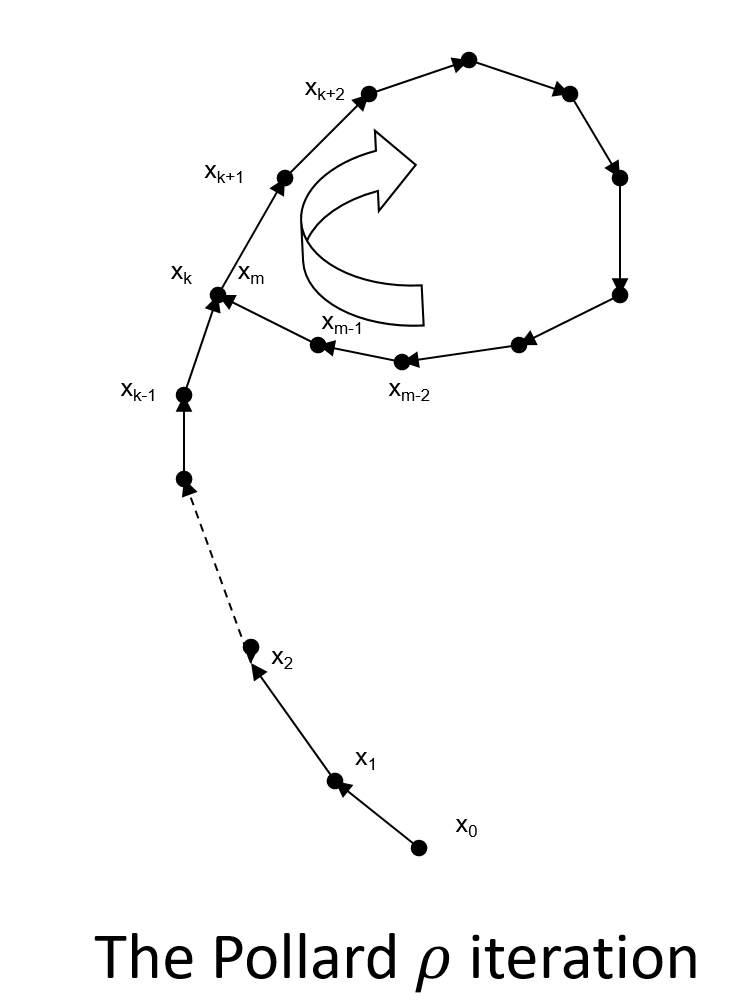}
     \caption{An illustration of the Pollard $\rho$ iterative walk.  Once the collision $x_m=x_k$ occurs, the walk enters into an infinite loop.  The shape of this diagram is the source of the moniker ``$\rho$''.   \label{fig:prho}}
   \end{figure}

\paragraph{\bf Algorithm A.2.2:~DLOGs via the Pollard $\rho$ walk.} Here the iteration $x_{n+1}$ is equal to one of three possibilities:~$x_n^2$, $x_n g$, or $x_n h$, with each possibility determined by the value of $x_n$ (but independently of $n$), and each occurring with probability   1/3.  This iteration is thus precisely (\ref{fkey}) written in multiplicative coordinates (and with the obvious adjustment to the meaning of $h$). Collisions again occur in time $O(\sqrt{\# G})$ \cite{KMPT} and are highly likely to be nondegenerate \cite{MV2}; all other details are identical to those in Algorithm A.2.1.\\

Note there is a subtle but important difference between these two algorithms:~Algorithm A.2.2 is superior in that at each step one requires only exponentiations to small powers.  It is thus significantly faster than Algorithm A.2.1, which requires  exponentiation to large powers and whose running time is therefore greater, in fact by  an additional multiplicative factor on the order of $\log(\# G)$.

\paragraph{\bf Algorithm A.2.3 (The Pollard $\rho$  \textbf{\textit{Algorithm}}):~Improvements using Floyd's cycle detection method.}
Algorithms A.2.1 and A.2.2 both require $O(\sqrt{N})$ storage in order to detect the first collision.  This is impractical for   large values of $N$ of   interest  in many applications.
The iterations in both algorithms enter  a loop once a collision occurs.  Note the following. If $x_k=x_n$ for some $k\le n$, then $x_{k+\ell}=x_{n+\ell}$ for any $\ell \ge 0$ because the iteration  is time-independent.  Thus $x_m=x_{m'}$ if $m\ge k$ and $m'\ge m$ is congruent to $m \pmod{n-k}$.  In particular, there exist $m\ge k$ such that $x_m=x_{2m}$, for example whenever $m$ is a multiple of $n-k$.  Thus Algorithm A.2.2 can be modified to use  much less storage by utilizing
 {\it
Floyd's  cycle detection} method, which simply checks if  $x_m=x_{2m}$. This requires only bounded storage, since it suffices to    run a duplicate copy of the iteration at half-speed.  The value of $m$ will typically be greater than that of the first collision time $n$.  The actual Pollard $\rho$ {\em algorithm} used in all practical implementations indeed uses Floyd's cycle detection algorithm -- it thus differs from Algorithm A.2.2 in that it may not detect the first collision.
\\

We have thus just seen that both Process 1 and Process 2 from \secref{sec:intro} naturally appear  in cryptographic attacks -- in fact, attacks on the same problem, viz., DLOG.  The underlying walks also have appeared in cryptosystems themselves.  The completely random ``birthday paradox'' walk is ubiquitous, since it occurs any time elements are selected uniformly at random from a set.
 The random walk in the Pollard $\rho$ walk is used in the stream cipher  \textsc{mv3}  \cite{KMMV}, and is closely modeled after the Pollard $\rho$ walk (see also Observation 2 at the end of \secref{sec:expander}).  Stream ciphers are fast random number generators used in cryptography for encrypting very large files, such as videos.  Collision time
 distribution enters  fundamentally into an aspect of their design.  A stream cipher outputs part of its internal state, which is updated at each iteration.  When collisions in the internal state (or at least parts of it) occur after short intervals, an attacker may be able to deduce secret information.  On the other hand,  unusually long collision times betray a highly non-random structure that an attacker might also be able to leverage.  It is thus desirable from this point of view to have collision times similar to those of the birthday problem collision times $B_N$  from the purely random walk in Process 1.  Thus verifying that normalized collision times match the Rayleigh distribution is a sign of good randomness properties of the internal workings of a stream cipher.

\subsection{Details of Figures~\ref{fig:ECCvsRSAnewa} and \ref{fig:ECCvsRSAnewb}}\label{sec:sub:figures}

\paragraph{\bf Details of Figure~\ref{fig:ECCvsRSAnewa}.}
The bottleneck in RSA key generation is the search for   primes  $p$ and $q$ in an interval $[X,\kappa X]$, where  $X\gg 1$ and $\kappa>1$.   The primes in our calculations were generated using the {\tt Pari/GP} number theory package \cite{parigp} as follows:~we first selected a uniformly random integer $r\in [X,\kappa X]$, and found the first prime $p>r$ using the {\tt nextprime}
command; the prime $q$ was generated in the same way.  The encryption exponent $e$ was chosen uniformly at random among the integers  which are coprime to $\phi(n)$ and lie in the interval $[1,\phi(n)]$.  The decryption exponent $d$ was then computed as the modular inverse of $e\pmod{\phi(n)}$.  Creating $e$ and $d$ is significantly faster than obtaining   $p$ and $q$. The actual CPU   timing to create the key was recorded by calling  the {\tt gettime()} command both immediately before the key is generated and again immediately after the key is  generated.

Each value of $X\gg 1$ and $\kappa>1$ gives rise to an ensemble ${\cal E}_{X,\kappa}$ of random RSA keys $(n,e)$.  With $\kappa$ fixed, as $X$ becomes large the histograms for the normalized timings for RSA key generation converge.  Moreover, one obtains the same histogram for different values of $\kappa$.  In this way RSA key generation timings give rise to a time-signature, displayed as the dotted curve in Figure~\ref{fig:ECCvsRSAnewa}.

Key generation for ECC  was made in the following way, using a randomly selected elliptic curve $E:y^2=x^3+x+r$ over $\F=\F_p$, where $p$ is a large random prime chosen in $[X,\kappa X]$ as above and $r$
is a random integer chosen uniformly from 1 to $p$.  Let $G$ denote the group of points of $E$ over $\F_p$.  Computations in $G$ were done using {\tt Pari/GP}:~the curves are first entered into  {\tt Pari/GP}  using the command {\tt E=ellinit([1,r],Mod(1,p))}.  Elliptic curves are more cryptographically secure if  $\# G$ is a prime not equal to $p$.  We thus discarded curves which had composite order.  A generator $g$ of $G$ was obtained using  the command {\tt g=ellgenerators(E)[1]}. Finally, the shared key $g^{xy}=(g^x)^y$ was created by selecting random integers $x$ and $y$ chosen uniformly in the range $[1,\#G]$ (using the commands {\tt x=random(ellcard(E))} and {\tt y=random(ellcard(E))}), and then using the command {\tt ellmul(E,ellmul(E,g,x),y)} to add $g$ to itself $xy$ times.
  The actual CPU time for key generation was again recorded using the {\tt gettime()} command.

 As above, each value of $X\gg 1 $ and $\kappa >1$ gives rise an ensemble ${\cal E}'_{X,\kappa}$ of random keys for ECC, i.e., for such $X$ and $\kappa$ we obtain a random elliptic curve $E$, conditioned to have prime order, and then from  $E$, a random key.  Again, with $\kappa$ fixed and $X$ large the histograms for the normalized key generation timings  converge  to a limit, which is independent of $\kappa$.  Thus ECC key generation timings give rise to a time-signature, displayed as the solid curve in  Figure~\ref{fig:ECCvsRSAnewa}.

The time-signature for RSA key generation was calculated by applying the normalization (\ref{normalizedtimesRay}) to 10,000 running times for generating keys with $n$ of size  $2^{4000}$; the resulting histogram for keys with $n$ of size  $2^{2000}$ was indistinguishable.  Likewise, the time-signatures for ECC key generation  were found from 10,000 normalized  running times to generate elliptic curves with $p$ of size $2^{70}$ (which is indistinguishable from the corresponding histograms   with $p$ of size $2^{50}$ or $2^{60}$).

\paragraph{\bf Details of Figure~\ref{fig:ECCvsRSAnewb}.}  Figure~\ref{fig:ECCvsRSAnewb} compares actual CPU running times vs. the number of iterative steps for the Pollard $\rho$ Algorithm A.2.3 to solve elliptic curve discrete logarithms.  We created instances of the discrete logarithm problem using {\tt Pari/GP} as follows. Given an elliptic curve  $E$ with generator $g$ entered using the commands {\tt E=} p{\tt ellinit([1,r],Mod(1,p))} and {\tt g=ellgenerators(E)[1]} as above,
and $x$ an integer in the range $[1,\#G]$,
the point $h=g^x$ was computed using the command {\tt h = ellmul(E,g,x)}.  In this way, we obtained a large data set to which the Pollard $\rho$ algorithm can be applied.

 For any given $h$ in our data set, the discrete logarithm of $h$ with respect to the generator $g$ was then computed  in {\tt Pari/GP} using the Pollard $\rho$ algorithm via the command {\tt elllog(E,h,g)}.  We used the {\tt gettime()} command as above to record the actual CPU time.  A straightforward   modification of the {\tt Pari/GP} source code also reports the step count.
  Working on the given elliptic curve $E$, we recorded both the actual CPU timing and step count of the Pollard $\rho$ algorithm for all the $h$'s in our data set.

The plots in   Figure~\ref{fig:ECCvsRSAnewb} display these timings.  They were
 first computed on a data set of  100,000 values of $h$ on a fixed elliptic curve $E$ having $p$ of size $2^{40}$.  This gives rise to two normalized histograms, one  for the actual CPU time (solid curve) and one for the step count (dashed curve).
  We repeated this experiment with a data set of 10,000 values of $h$ on a different elliptic curve  $E'$ having $p$ of size $2^{45}$.  This gave rise to two additional normalized histograms, which agreed perfectly with their respective counterparts (i.e., solid-to-solid and dashed-to-dashed) from $E$.  From these computations we see  in addition that the solid and dashed curves are very close to each other; that is, there is little difference in measuring the time-signature of the Pollard $\rho$ algorithm by actual CPU running times as opposed to step count.


\begin{thebibliography}{99}

\bibitem{CP} Michael Camarri and Jim Pitman, {\it Limit distributions and random trees
derived from the birthday problem
with unequal probabilities}, Electronic Journal of Probability {\bf 5}, 1--18 (2000).
\newline\url{https://projecteuclid.org/download/pdf_1/euclid.ejp/1457376437}

\bibitem{eigencite1} Percy A.~Deift, Govind Menon, Sheehan Olver and Thomas Trogdon, {\it Universality in numerical computations with random data}, Proc. Natl. Acad. Sci. U. S. A. {\bf 111}, 14973--14978 (2014).

\bibitem{eigencite3} Percy Deift and Thomas Trogdon, {\it Universality for eigenvalue algorithms on sample covariance matrices}, SIAM J. Num. Anal. {\bf 55}, 2835--2862 (2017).


\bibitem{eigencite2} Percy Deift and Thomas Trogdon, {\it Universality for the Toda Algorithm to Compute the Largest Eigenvalue of a Random Matrix}, Commun. Pure Appl. Math. {\bf 71}, 505--536 (2017).

\bibitem{DH} Whitfield Diffie and Martin E. Hellman, {\it New directions in cryptography}, IEEE Transactions on Information Theory {\bf 22} (6), 644--654 (1976).


\bibitem{keeloq}Thomas Eisenbarth, Timo Kasper, Amir Moradi, Christof Paar, Mahmoud Salmasizadeh, Mohammad T. Manzuri Shalmani, {\it On the Power of Power Analysis in the Real World: A Complete Break of the KeeLoq Code Hopping Scheme}, in {\it Advances in Cryptology -- CRYPTO 2008}, Springer Berlin Heidelberg, pp.203--220.
 \url{https://doi.org/10.1007/978-3-540-85174-5_12}



\bibitem{KMMV} Nathan Keller, Stephen D.~Miller, Ilya Mironov, and Ramarathnam Venkatesan, {\it MV3: A new word based stream cipher using
rapid mixing and revolving buffers}, in {\it Topics in Cryptology -- Procedings of CT-RSA 2007}, Springer Verlag Lecture Notes in Computer Science, {\bf 4377} (2007),   1--19.


\bibitem{KMPT} Jeong-Han Kim, Ravi Montenegro, Yuval Peres, and Prasad Tetali, {\it A Birthday Paradox for Markov chains with an optimal bound for collision in the Pollard Rho algorithm for discrete logarithm},  	Annals of Applied Probability, {\bf 20}, No. 2, 495--521 (2010).

    \bibitem{Mo}  Jeong-Han Kim, Ravi Montenegro, and Prasad Tetali, {\it Near Optimal Bounds for Collision in Pollard Rho
for Discrete Log}, in
Proceedings of the 48th Annual IEEE Symposium on Foundations of
Computer Science,  215-–223. Washington, DC: IEEE, 2007.

\bibitem{Kolm} A.N.~Kolmogorov, {\it Sulla determinazione empirica di una legge di distribuzione}. Giornale dell'Istituto Italiano degli Attuari, Vol. {\bf 4} (1933), pp. 83--91.


\bibitem{Lilliefors} H.~Lilliefors,  {\it On the Kolmogorov--Smirnov test for normality with mean and variance unknown}, Journal of the American Statistical Association, {\bf 62}, pp. 399--402 (1967).

\bibitem{MV1} Stephen D.~Miller and Ramarathnam Venkatesan, {\it Spectral analysis of Pollard rho collisions},  Algorithmic number theory, 573--581, Lecture Notes in Comput. Sci., {\bf  4076}, Springer, Berlin, 2006.


\bibitem{MV2} Stephen D.~Miller and Ramarathnam Venkatesan, {\it Non-degeneracy of Pollard rho collisions}, Int. Math. Res. Not. IMRN {\bf 2009}, no. 1,  1--10


\bibitem{parigp} Pari/GP computer algebra system, \url{https://pari.math.u-bordeaux.fr/}.


\bibitem{eigencite4} Christian W. Pfrang, Percy Deift and Govind Menon, {\it How long does it take to compute the eigenvalues of a random symmetric matrix?}, in {\it Random matrix theory, Interact. Part. Syst. Integr. Syst.}, MSRI Publ. {\bf 65}, 411--442 (2014).

\bibitem{Pollard} John M.~Pollard, {\it
Monte Carlo methods  for index computation$\pmod p$},
 Mathematics of
Computation
{\bf 32}
(1978), no. 143, 918--924.

\bibitem{rabin} Michael O. Rabin,
  {\it Probabilistic algorithm for testing primality}, J. Number Theory {\bf 12}, 128--138 (1980).

\bibitem{shoup} Victor Shoup,
Lower  bounds  for  discrete  logarithms  and  related  problems, Advances in Cryptology--EUROCRYPT ’97 (Konstanz), Lecture Notes in Comput. Sci., {\bf 1233}, Springer, Berlin, 1997, pp. 256–266,
Updated version at
http://www.shoup.net/papers/dlbounds1.pdf

\bibitem{T} Edlyn Teske, {\it On random walks for Pollard's rho method}, Mathematics of Computation
{\bf 70}, Number 234,  809--825 (2000).


\end{thebibliography}
\end{document}